\documentclass[lettersize,journal]{IEEEtran}
\usepackage{amsmath,amsfonts}
\usepackage{algorithmic}
\usepackage{algorithm}
\usepackage{array}
\usepackage[caption=false,font=normalsize,labelfont=sf,textfont=sf]{subfig}
\usepackage{textcomp}
\usepackage{stfloats}
\usepackage{url}
\usepackage{verbatim}
\usepackage{graphicx}
\usepackage{cite}
\usepackage{color}
\begin{document}

\title{Localization in Multipath Environments via Active Sensing With Reconfigurable Intelligent Surfaces}

\author{Yinghan Li,~\IEEEmembership{Graduate Student Member,~IEEE}, and Wei Yu,~\IEEEmembership{Fellow,~IEEE}
\thanks{Manuscript accepted in \emph{IEEE Communications Letters}. The authors are with The Edward S.\ Rogers Sr.\ Department of
Electrical and Computer Engineering, University of Toronto, Canada. (e-mails: yinghan.li@mail.utoronto.ca, weiyu@ece.utoronto.ca). 
This work is supported 
by Huawei Technologies Canada. 
}

}



\maketitle

\thispagestyle{empty}

\begin{abstract}
This letter investigates an uplink pilot-based wireless indoor localization
problem in a multipath environment for a single-input single-output
(SISO) narrowband communication system aided by reconfigurable intelligent
surface (RIS). The indoor localization problem is challenging because the
uplink channel consists of multiple overlapping propagation paths with varying
amplitudes and phases, which 
are not easy to differentiate. This letter proposes the use of RIS 
capable of adaptively changing its reflection pattern to
sense such a multiple-path environment. Toward this end, we train a
long-short-term-memory (LSTM) based controller to perform adaptive sequential
reconfigurations of the RIS over multiple stages and propose to group
multiple pilots as input in each stage. Information from the multiple paths is
captured by training the LSTM to generate multiple RIS configurations to align
to the different paths within each stage.  Simulation results show that the
proposed approach is effective in significantly reducing training complexity 
while maintaining localization performance at fixed number of pilots.
\end{abstract}


\section{Introduction}

 Localization has garnered substantial attention in the field of wireless 
communication due to its relevance in many applications, including navigation, 
mapping, target tracking, robotics, etc \cite{bourdoux20206g}.  This letter
treats the uplink localization problem in which a user equipment (UE) transmits
known pilots over the air, and an access point (AP) aims to determine the
location of the UE based on the received pilots. As the wireless channel is a
function of the user's location, conceptually, uplink localization is
a problem of inverse mapping---from the wireless channel realization
(which can be estimated from the received pilots) to the user location. 

The uplink localization problem is challenging in a narrowband system,
especially when the UE and the AP are equipped with only a single antenna. 
In this case, the AP cannot obtain angular information of the incoming signal
using array signal processing techniques. The situation is further exacerbated in a
multipath indoor environment, where the pilot signal from the UE is reflected off the
surrounding environment over multiple paths (possibly multiple times) before
arriving at the AP. In this case, the multiple propagation paths with varying
amplitudes and phases inevitably overlap, making them difficult to differentiate.

This letter explores the use of reconfigurable intelligent surface (RIS) to
aid the localization task. 
The idea is that by adaptively
reconfiguring the reflection pattern of the RIS over multiple stages, 
the AP would be able to ``scan'' the environment through the RIS, thereby 
achieving accurate localization. 
 
 
 
To this end, this letter considers a narrowband single-input single-output
(SISO) system, in which the UE is equipped with an isotropic antenna that
periodically broadcasts pilots in all directions. The pilots undergo
reflections both by the RIS and by various obstacles in the environment, before
ultimately arriving at the AP. The AP can adaptively adjust the reflection
coefficients of the RIS elements sequentially in order to enhance the
localization accuracy. 
At the end, based on the received pilots across all stages, the AP can
produce an estimate of the position of the UE. 

The main question is how to adaptively configure the RIS reflection coefficients 
across the multiple pilot stages to achieve the best localization accuracy.  
Observe that because the localization process is over multiple stages, the AP should 
be able to configure the RIS coefficients based on what it has already learned 
about the UE location so far. Thus, the design of the optimal RIS configuration is 
in effect an optimization over the space of functional mappings from the observations
so far to the RIS coefficients. The design of this so-called \emph{active sensing} 
strategy is highly nontrivial, because of the high dimensionality of the optimization spaces.

In this end, prior works have proposed a \emph{data-driven} approach for active
sensing in the contexts of beam alignment \cite{Jiang2023} and localization
\cite{Zhang2023} using a long short-term memory (LSTM) structure that effectively summarizes historical observations into a state vector to derive subsequent RIS configurations.
However, the work \cite{Zhang2023} primarily focuses on channel models where the AP and the UE only have one line-of-sight (LoS) path and one
single non-line-of-sight (NLoS) path through the RIS. 
In an indoor environment, capturing information from multiple
paths is essential for improving localization accuracy. When the LSTM
structure of \cite{Zhang2023} is applied to the multipath scenario, it would
require a much longer LSTM chain than the single-path case. This leads to longer
pilots as well as increased training and inference complexity.

The main contribution of this letter is to address the crucial question of how to design
the neural network architecture to suit the active sensing task in a multipath environment.
Toward this end, we modify the LSTM structure in \cite{Zhang2023}, where each
LSTM stage corresponds to only a single pilot. Instead, we divide the pilots
into groups,then group multiple received pilots as input to each LSTM stage.
This allows a well-trained LSTM to accumulate a variety of path information,
based on the multiple pilots in each stage.  The results demonstrate that the
proposed approach can significantly reduce the training and inference
complexity while ensuring excellent localization performance. In essence, the
proposed strategy is able to effectively track multiple reflected paths at each
stage through adaptive reconfigurations of the RIS, thus allowing it to provide
valuable information to the subsequent stages.   
 
Prior works on RIS-aided localization generally fall into two categories:
model-based and model-free methods.  In the model-based category, studies such as 
\cite{liu2021reconfigurable,Elzanaty21,Wymeersch20, Fascista22,Keykhosravi21} use
analytical approaches to optimize the RIS configuration for enhancing location 
accuracy, for example, by setting the objective as minimizing the Cramér-Rao 
lower bound (CRLB). 

These model-based methodologies 
typically rely on the functional relationships between the user location and 
the time-of-arrival (ToA) or angle-of-arrival (AoA) to derive a Fisher information
matrix (FIM) for the parameter of interest (i.e., the user location). But this approach becomes impractical
in multipath scenarios, because distinguishing ToAs and AoAs for different
paths in a narrowband indoor environment is difficult. Even if multiple paths
can be distinguished, relating the user locations to the ToAs or AoAs of the 
multiple paths (which are reflected by unknown surfaces) is an extremely 
challenging task. 
On the other hand, model-free approaches have become increasingly important with the rapid advances in machine learning.   
In \cite{nguyen2021wireless}, the authors propose a
fingerprinting method that maps the received signal strengths (RSS) at various RIS configurations to the locations. Supervised learning is
utilized to choose a subset of multiple configurations, with an aim to streamline the
feature vector and to reduce the complexity of the matching process.  Moreover
in \cite{Wang23}, deep reinforcement learning (DRL) is utilized to optimize the RIS
configurations, with an aim to improve localization accuracy for the RIS-aided
fingerprinting-based localization problem.  
However, the model-free approaches mentioned
above all utilize predefined RIS configurations without actively reconfiguring
them\cite{nguyen2021wireless, Wang23}, overlooking the potential
for active reconfiguration to further enhance dynamic performance. This letter
follows the work of \cite{Zhang2023} in using an LSTM-based approach for actively
designing the RIS configuration to better localize a remote user. We extend
the work of \cite{Zhang2023} in incorporating the indoor multipath scenario, 
which has not been considered previously.

\section{System Model And Problem Formulation}

\subsection{System Model}

\begin{figure}[t]
\centerline{\includegraphics[width=0.8 \linewidth]{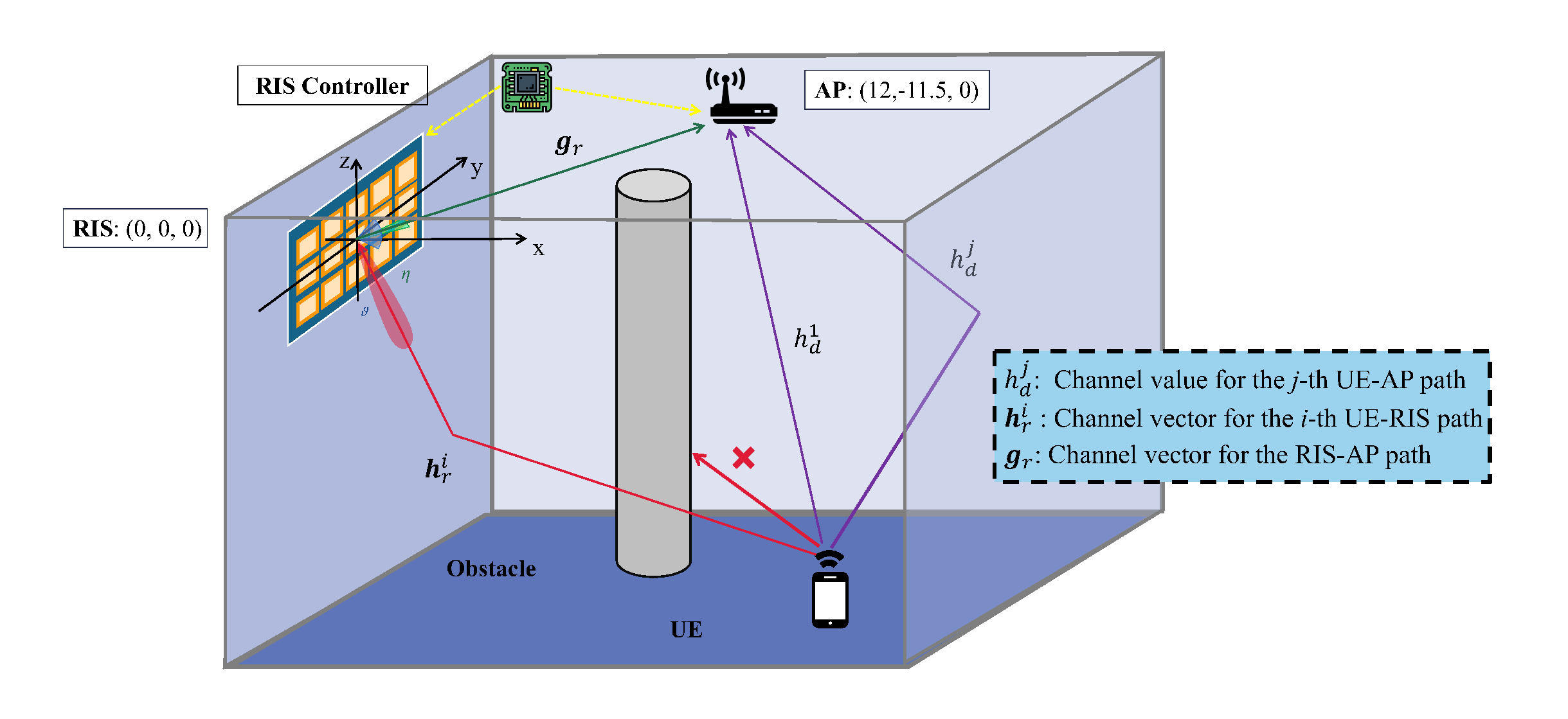}}
\caption{{RIS-assisted localization with multipath.}}
\label{fig1}
\end{figure}

Consider a wireless uplink indoor localization task in an environment
comprising a single-antenna UE, a single-antenna AP, and a RIS.
The AP and RIS are deployed as shown in Fig.~\ref{fig1}.
 The AP possesses the capability to
adjust received pilots by tuning the phase shifts of RIS elements through an
RIS controller.  The UE transmits pilots isotropically over the air. 
Signals from the UE undergo multiple reflections, while the direct path to the RIS is blocked.
We assume that a transmitted signal may undergo at most $R$ reflections.
Paths with more than $R$ reflections are excluded from the analysis due to
energy loss resulting from the multiple reflections.

In this letter, we adopt quasi-static block fading channels, where the channels are assumed to remain stationary across the stages. 
The number of stages is defined as $T$, and during each of these stages, $L$ consecutive pilots are transmitted. 
The total number of elements in the RIS is denoted as $M$.
For the $l$-th uplink pilot transmission in the $t$-th stage, the reflection coefficients of RIS are represented by $\boldsymbol{\theta}_l^{(t)} = [e^{j \delta^{t}_{l,1}}, e^{j \delta^{t}_{l,2}}, \cdots, e^{j \delta^{t}_{l,M}}]^{\top} \in \mathbb{C}^{M}$, where ${\delta^{t}_{l,m}} \in [0, 2\pi)$ represents the phase shift of the $m$-th RIS element.
Let $N_d $ represent the number of paths from the UE to the AP.
The UE-AP channel across multiple paths is expressed as $\mathbf{h}_d=[h_d^1,h_d^2,\cdots,h_d^{N_d}]^{\top}$, where $h_d^i$ represents the $i$-th UE-AP path channel. 
Let $N_r $ represent the number of paths from the UE to the RIS.
The UE-RIS channel is denoted as $\mathbf{h}_r=[\mathbf{h}_r^1,\mathbf{h}_r^2,\cdots,\mathbf{h}_r^{N_r}]$, 
where $\mathbf{h}_r^{i} \in \mathbb{C}^{M}$ defines the channel vector for the $i$-th UE-RIS path, associated with the $M$ elements of the RIS.
Note that the dimensions of $\mathbf{h}_d $ and $\mathbf{h}_r$ vary based on the number of reflection paths $N_d, N_r$, according to the specific scenario.
The channel models for the $i$-th path are given as follows:
\begin{align}
    h_d^i &= A_d^i e^{j\Phi_d^i}, \\
    \mathbf{h}_r^i &= A_r^i e^{j\Phi_r^i} \mathbf{a}(\phi^i, \psi^i),
\end{align}
where $ A_d^i $ and $ A_r^i $ are the amplitude gains of the $ i $-th UE-AP and UE-RIS paths, respectively; $ \Phi_d^i $ and $ \Phi_r^i $ represent the corresponding phase shifts, $\phi^{i} $ and $\psi^{i} $ denote the azimuth and elevation angles of the $ i $-th path from the UE to the RIS. The spatial phase shift vector $\boldsymbol{a}(\phi^{i}, \psi^{i})$ of the $n$-th element, associated with the angles $\phi^{i}$ and $\psi^{i}$, is given by:
\begin{equation}
[\boldsymbol{a}(\phi^i, \psi^i)]_n = e^{ \frac{2j\pi d_{\mathrm{R}}}{\lambda_c} \left( v_1(n, N_c) \sin (\phi^i) \cos (\psi^i) + v_2(n, N_c) \sin(\psi^i) \right) },
\end{equation}
where $d_{\mathrm{R}}$ is the distance between two reflective elements of the RIS, $\lambda_c$ is the carrier frequency, and $v_1(n, N_c)=\bmod (n-1, N_c)$ and $v_2(n, N_c)=	\left\lfloor\frac{n-1}{N_c}\right\rfloor$. Here, $N_c$ is the number of columns of the RIS. We use $\mathbf{g}_{{r}}\in \mathbb{C}^{M}$ to describe the channel from the $M$ RIS elements to the AP. The channel for the RIS-AP path is represented as follows:
\begin{equation} 
{\boldsymbol{g}}_{\mathrm{r}}=A_g e^{j \Phi_g} \boldsymbol{a}\left(\eta, \vartheta\right)^{\mathrm{H}},
\end{equation}
where $ A_g $ and $ \Phi_g $ represent the amplitude gains and phase shifts of the RIS-AP path at the AP, and $\eta$ and $\vartheta$ denote the azimuth and elevation angles of the RIS-AP path.
In a narrowband system, where the time delays in signal transmission across different paths are negligible, we assume that pilots transmitted from multiple paths arrive at the AP simultaneously. 
Consequently, the $l$-th received pilot at the AP during the $t$-th stage, denoted as $\hat{y}_l^{(t)}$, is expressed as:
\begin{equation}
\hat{y}_l^{(t)} = \sqrt{P_u}\left(\mathbf{h}_d^{\top}\textbf{1} + \mathbf{v}_{{r}}^{\top} \boldsymbol{\theta}_l^{(t)}\right)x_l^{(t)} + n_l^{(t)},
\end{equation}
where $P_u$ represents the transmission power, $\mathbf{v}_{{r}}=\operatorname{diag}(\mathbf{h}_{{r}}\mathbf{1}) \mathbf{g}_{{r}}\in \mathbb{C}^{M}$ represents the cascaded channel between the BS and the UE through the reflection at the RIS, ${x}_l^{(t)}$ represents the $l$-th transmitted pilot in the $t$-th stage and  ${n}_l^{(t)} \sim \mathcal{C} \mathcal{N}\left(0, \sigma_0^2\right)$ is the additive white Gaussian noise at the AP associated with ${x}_l^{(t)}$. Here, $\textbf{1}$ represents an all-one vector.

\subsection{Problem Formulation}
\label{subsec:Problem}
\begin{figure*}[tbp]
\centerline{\includegraphics[width=0.73 \linewidth]{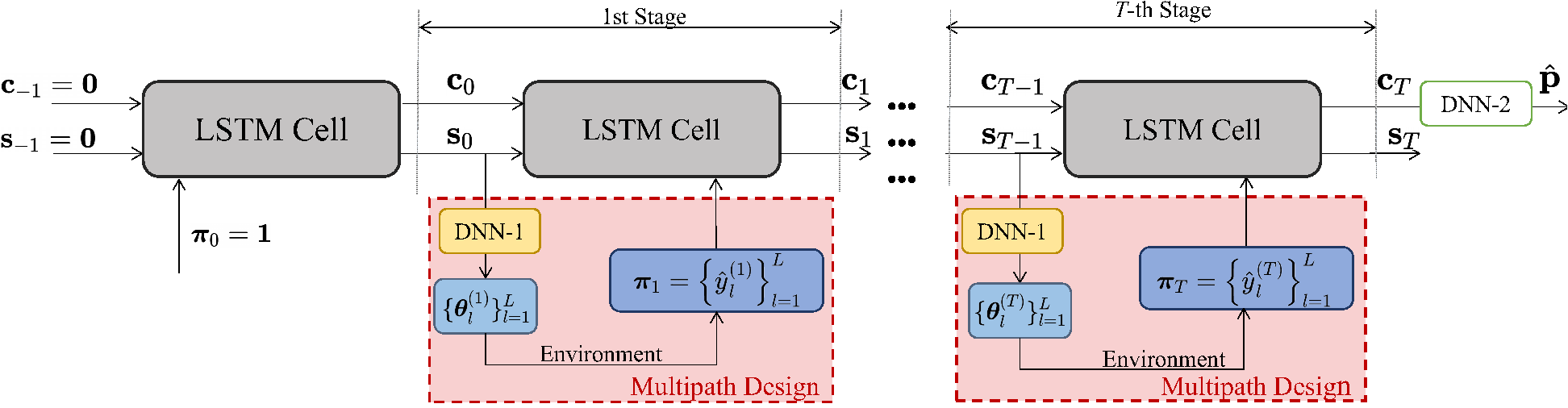}}
\caption{Proposed LSTM with $T$ concatenated units corresponding to $T$ stages. Each stage takes $L$ pilots as inputs.}
\label{LSTM}
\end{figure*}
While prior studies have shown that accurate positioning is achievable in multipath conditions with large antenna arrays on both APs and UEs \cite{kim20205g}, or through ultra-wideband (UWB) measurements \cite{6655524}, the focus of this letter is different. 
In this study, we propose an approach to address the localization problem in
multipath SISO narrowband scenarios, while utilizing the reconfigurable RIS. 
Our main innovation is in grouping multiple pilots as input, thus allowing
the neural network structure to focus on the multiple paths.

Building upon the proposed grouping approach, the central problem addressed in this letter is the estimation of the UE's location by leveraging the grouped pilots over $T$ stages, with each stage receiving $L$ pilots.
In the $t$-th stage, the configuration of the RIS, denoted as $\{\boldsymbol{\theta}_l^{(t)}\}_{l=1}^L$,  is adjusted based on \emph{groups} of received pilots in the preceding $t-1$ stages $\{\{\hat{y}_l^{(\tau)}\}_{l=1}^L\}_{\tau=0}^{t-1}$. 
Define the mapping from the received pilots to the RIS configuration for stage $t$ as 
\begin{equation} 
\{\boldsymbol{\theta}_l^{(t)}\}_{l=1}^L =\mathcal{G}^t\left(\{\{\hat{y}_l^{(\tau)}\}_{l=1}^L\}_{\tau=0}^{t-1}\right).
\end{equation}
Define the mapping from the received pilots to 
the final estimated location $\hat{\mathbf{p}}$ as
\begin{equation} 
\hat{\mathbf{p}}=\mathcal{F}\left(\{\{\hat{y}_l^{(\tau)}\}_{l=1}^L\}_{\tau=0}^{T}\right) .
\end{equation}
The RIS-aided localization problem is that of minimizing the error between the estimated location $\hat{\mathbf{p}}$ and the actual location $\mathbf{p}$ through received pilots, and can be formulated as follows:
\begin{equation} 
\hspace{-5mm}
\begin{array}{cl}
\underset{ \left\{ \mathcal{G}^t(\cdot)\right\} _{t=0}^{T-1}, \mathcal{F}(\cdot)}{\operatorname{minimize}} & \mathbb{E}\left[\|\hat{\mathbf{p}}-\mathbf{p}\|_2^2\right] \\
\text { subject to } & \left|[\boldsymbol{\theta}_{l}^{(t)}]_n\right|=1, \forall  n,t, l,\\
&\{\boldsymbol{\theta}_l^{(t)}\}_{l=1}^L =\mathcal{G}^t\left(\{\{\hat{y}_l^{(\tau)}\}_{l=1}^L\}_{\tau=0}^{t-1}\right),\forall  t,\\
& \hat{\mathbf{p}}=\mathcal{F}\left(\{\{\hat{y}_l^{(\tau)}\}_{l=1}^L\}_{\tau=0}^{T}\right),
\end{array}
\end{equation}
where $[\boldsymbol{\theta}_{i}^{(t)}]_n$ represents the $n$-th element of the vector $\boldsymbol{\theta}_{i}^{(t)}$.
Due to the complex nature of the functions \( \mathcal{G}^t_i \) and \( \mathcal{F} \), finding a global optimal solution analytically for this problem is intractable. 
To make this problem tractable, this letter utilizes a data-driven approach based on LSTM.
\section{Neural Network Architecture for Multipath Scenarios}

Our proposed method employs LSTM to estimate UE locations by utilizing the superimposed received power and phases of transmitted pilots from multiple paths. 
To manage the complexity arising from capturing multiple paths, we suggest grouping several pilots together as input at each stage. 
The proposed method increases the input dimension of the LSTM, thereby enhancing its capacity to acquire information from multiple paths in each stage.

As shown in Fig.~\ref{LSTM}, the proposed architecture comprises three main components. The first component is the LSTM neural network, where the number of LSTM cells corresponds to the number of pilot stages.
The cell state $\mathbf{c}_t$ serves as memory unit, retaining information from prior stages, including the different paths. 
The state vector $\mathbf{s}_t$ retains essential information for setting the subsequent RIS configurations.

The second component is the multipath design segment. Here the input $\boldsymbol{\pi}_t$, representing a \emph{group} of $L$ received pilots at the $t$-th stage, is used to update the state vector.
Further, a fully connected neural network (FCNN), denoted as DNN-1, takes the LSTM-derived state vector as input and dynamically updates the configuration of the RIS for the subsequent stage.
 
The third component, denoted as DNN-2, is a FCNN for estimating the final location of the UE.
 
The key idea here is to group pilots into
$\boldsymbol{\pi}_t$, thus making it possible for the LSTM to incorporate up to $L$ paths in the environment. In this letter, $\boldsymbol{\pi}_t$ is defined as $\left\{\Re\left(\{\hat{y}_l^{(t)}\}_{l=1}^L\right),\Im\left(\{\hat{y}_l^{(t)}\}_{l=1}^L \right)\right\}$, where $\Re $ and $\Im$ respectively denote the real and imaginary parts of $L$ received pilots at $t$-th stage.

Using the known UE location as the label, we adopt an end-to-end learning approach. The loss function is formulated as the average mean squared error (MSE) of the estimated location $\hat{\mathbf{p}}$. It is expressed as  $\dfrac{1}{{M_{tr}}} \sum_{m=1}^{M_{tr}}\left\|\hat{\mathbf{p}}-\mathbf{p}\right\|_2^2$, where $M_{tr}$ represents the number of training samples.

\section{Performance Evaluation and Interpretation }

 \begin{figure}[tbp]
\centerline{\includegraphics[width=0.97 \linewidth]{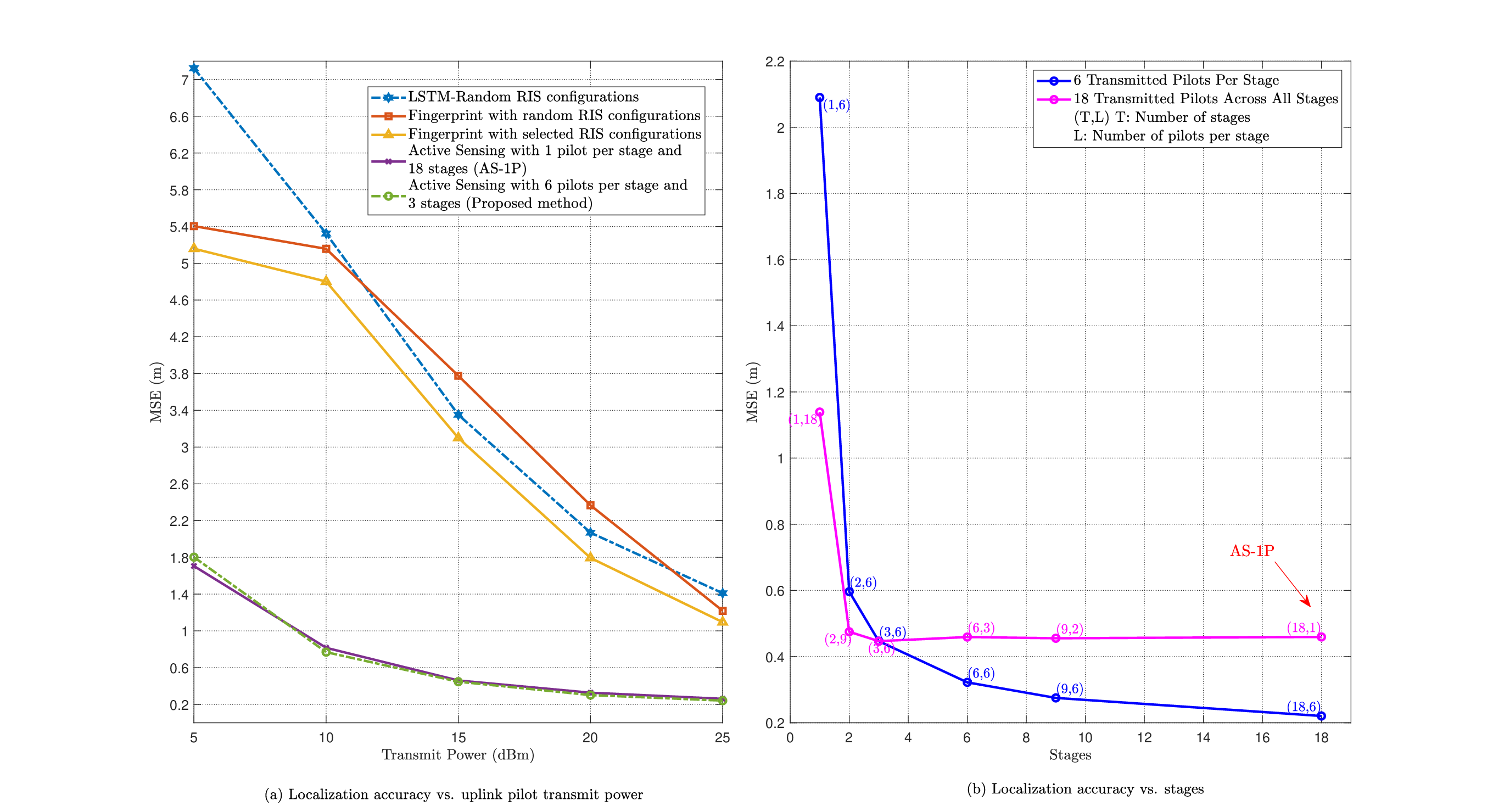}}
\caption{Localization accuracy performance of the proposed method.}
\label{accuracy}
\end{figure}

\begin{figure}[tbp]
\centerline{\includegraphics[width=0.97 \linewidth]{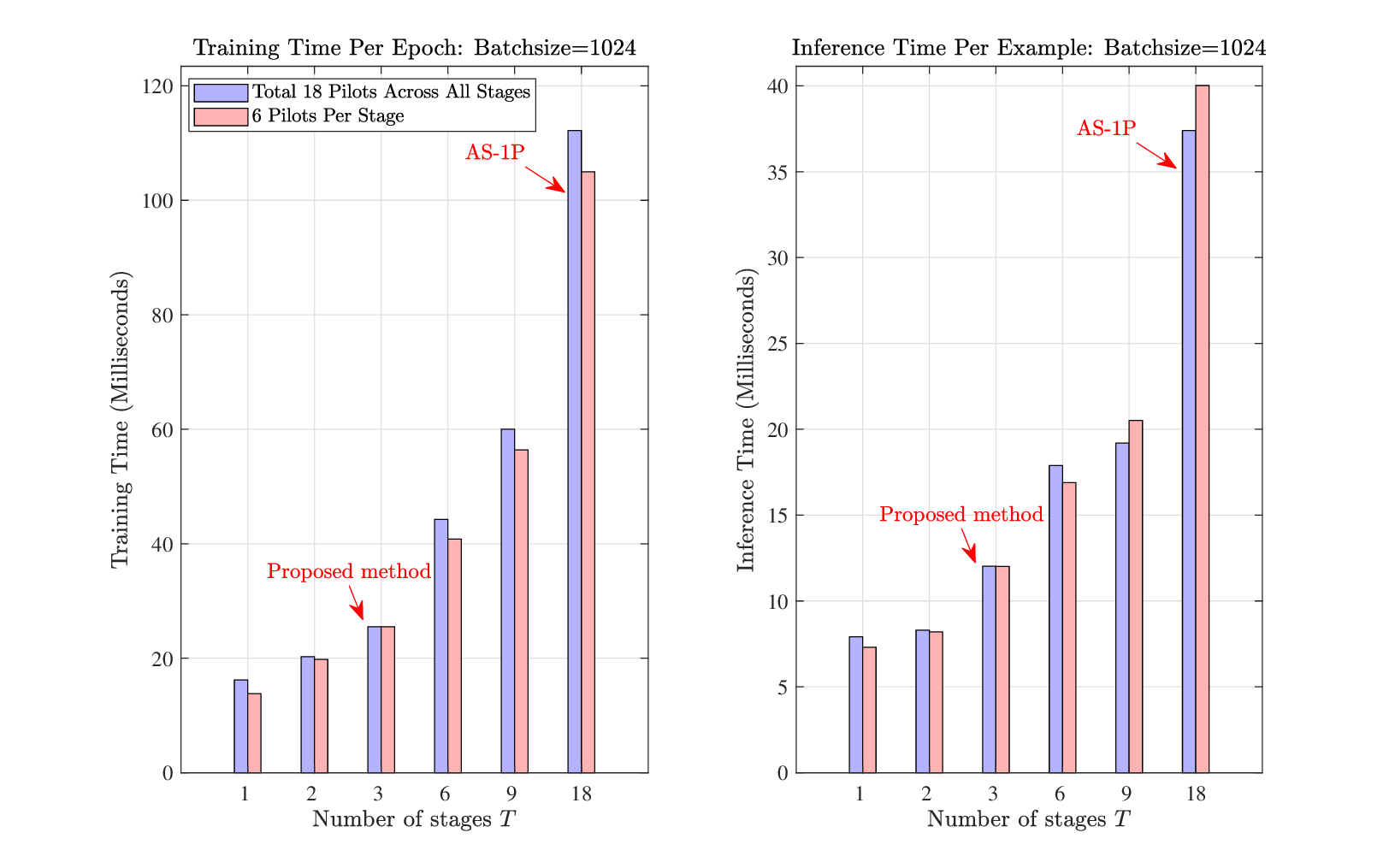}}
\caption{ Training time and inference time vs. different number of stages $T$.}
\label{time}
\end{figure}

\begin{figure*}[tbp]
\centerline{\includegraphics[width=0.780 \linewidth]{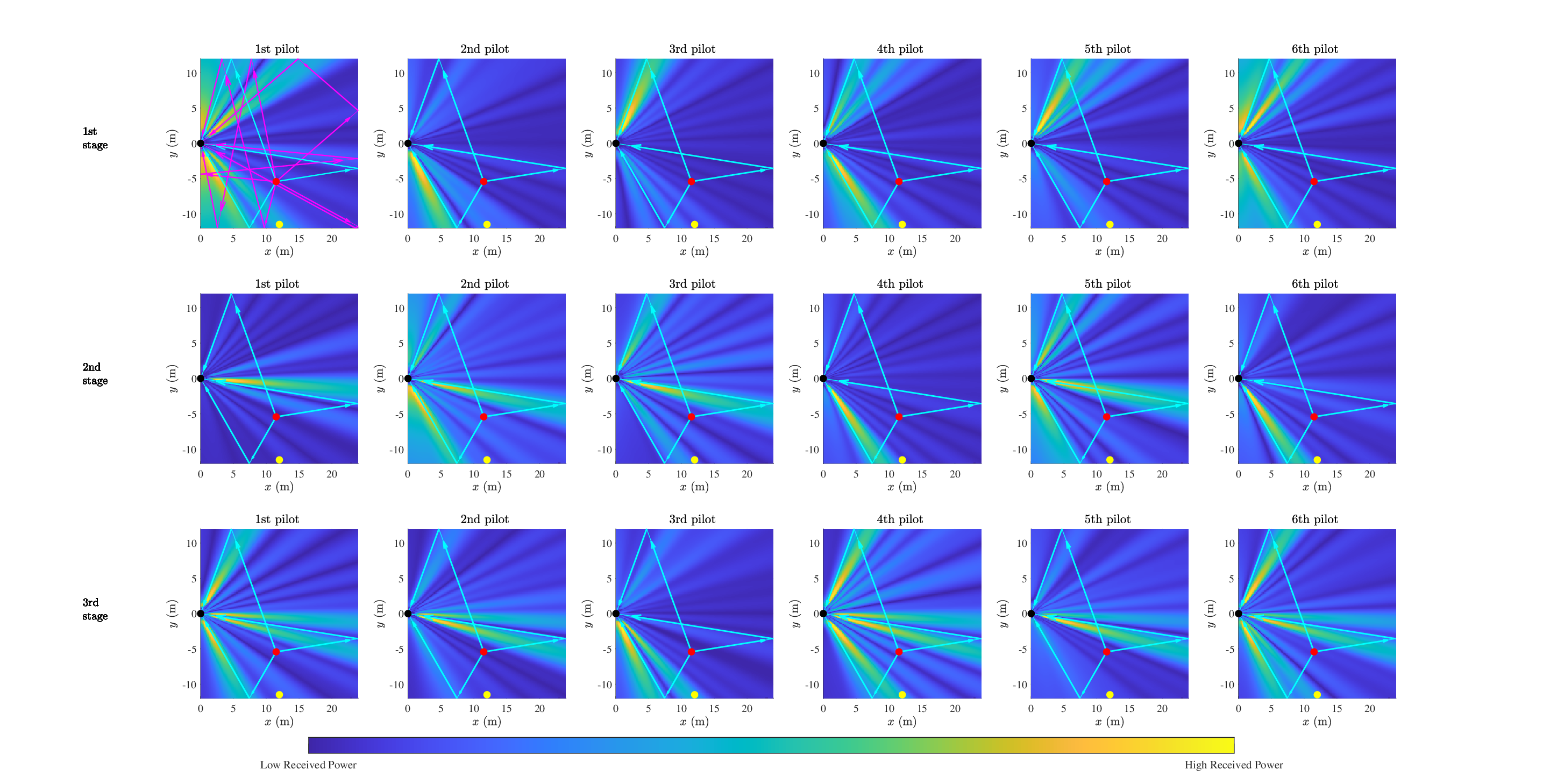}}
\caption{Received signal strength over one UE-RIS-AP path with varying start points across the (x, y)-plane under different RIS configurations ($T=3$, $L=6$). The configurations of RIS are obtained from the trained LSTM based on the proposed method.}
\label{response}
\end{figure*}

The simulation setup, as shown in Fig.~\ref{fig1}, involves a room measuring $24$ meters in both length and width, and $3$ meters in height.
The complex permittivity and conductivity of the four reflective walls in the room are set as $2.5 $ $-$ $ 0.3j$ and $0.03$ respectively, while the ceiling and floor are constructed from non-reflective materials.
The AP and the RIS are positioned at Cartesian coordinates $[12,-11.5,0 ]$ and $[0,0,0]$, respectively, in meters.
UEs are randomly distributed within a square region bounded by $[4, 24]$ on the x-axis and $[-8, 12]$ on the y-axis, with a constant z-coordinate of $-0.5$ meters. The RIS, mounted on a wall within the (y, z)-plane, comprises a $16 \times 4$ uniform rectangular array of 64 elements.

We limit analysis to paths that undergo fewer than $R=5$ reflections. We set the transmission power at $15$ dBm, center frequency at $5.8 \times 10^9$ Hz, and noise power at $\sigma_{0}^{2} = -100$ dBm. 
The RSS for different locations is generated based on \cite{lyz22, Neeraj18}.  A total of $10^6$, $10^3$ and $10^4$  samples have been generated for training, validation, and inference, respectively. The number of iterations is set at $5\times 10^5$. The batch size for every iteration is $1024$. 

\subsection{Baseline Schemes}
{ \subsubsection{Fingerprint-Based Methods \cite{nguyen2021wireless}} 
The fingerprint-based methods rely on a database which maps each $0.1\,\text{m} \times 0.1\,\text{m}$ block within the area of interest to a corresponding vector of RSS's at the AP. Two RIS configuration methods are considered: i) The reflective coefficients of the RIS are randomly configured; ii) The sequence of RIS configurations is selected from 1000 randomly generated sets. The selection criterion aims to maximize some dissimilarity function, e.g., as proposed in \cite{nguyen2021wireless}.

 \subsubsection{LSTM With Random RIS Configurations} The RIS is randomly configured rather than learned. A LSTM is used to map the RSS's over all stages to the estimated location. To ensure a fair comparison, the number of trainable parameters of the LSTM is chosen to be the same as in the proposed method.
  
 \subsubsection{Active Sensing With One Pilot per Stage (AS-1P) \cite{Zhang2023}}  The RIS is reconfigured immediately upon the AP receiving
one pilot signal. This is equivalent to transmitting a single pilot at
each stage, i.e., setting $ L = 1 $. To ensure fairness, we compare schemes 
with the same total number of pilots, e.g., AS-1P over 18 stages versus the
proposed method with 6 pilots per stage over 3 stages. 
Again, the number of trainable parameters in the LSTM for AS-1P is kept the same as in the proposed method.}

\subsection{Simulation Results}
 We first examine the localization performance versus transmit power levels in Fig.~\ref{accuracy}(a). It has been seen Fig.~\ref{accuracy}(a) that the active sensing-based methods (i.e., AS-1P and the proposed method) achieve higher localization accuracy across different transmit power levels as compared to the other benchmarks. This demonstrates that adaptively designed RIS based on historical RSS can improve the accuracy of location estimation.
 The localization performances for various settings of the number of stages $T$ and
number of pilots $L$ are shown in 
Fig.~\ref{accuracy}(b). 
It is observed that the performance significantly improves either if we let $L$
increase while fixing $T$, or let $T$ increase while fixing $L$. In particular, if we fix the number of stage $T=18$,
having 6 pilots significantly reduces the localization error (by more than 50\%) 
compared to one pilot per stage. The same is true for $T=6$ or $T=9$.

On the other hand, we note that localization performance is mostly a function of the total 
number of pilots across all the stages and not the number of pilots per stage.  
The performance for $(T,L) = (3,6)$ is almost identical to $(6,3)$, $(9,2)$, or $(18,1)$.
The key advantage of grouping multiple pilots per stage is actually in the
significant LSTM training and inference complexity reduction due to grouping.
In Fig.~\ref{time}, we assess complexity by measuring the 
real execution time on a computer with NVIDIA GeForce GTX 1080 Ti GPU. 
As illustrated in Fig.~\ref{time}, the complexity is almost exclusively a function of 
the number of sensing stages $T$; the number of pilots per stage $L$ has minimal impact 
on the execution time. 
The fact that increasing dimension per stage does not influence the execution time is due to the
inherent parallel computing capabilities of the GPU.  Thus, with the same
total number of pilots, the proposed method, which groups multiple pilots per stage,
thereby reducing the number of stages, can exhibit a notable reduction in execution time,
as compared to AS-1P without sacrificing performance. 
At the total number of pilots of 18, the sweet operating spot is at $T=2$ or $3$ with $L=9$ or $6$ pilots per stage, which 
results in 5-fold or 4-fold complexity saving as compared to $T=18$ and $L=1$. 

In other words, as comparing to $(L,T)=(18,1)$ scheme of AS-1P, we can 
reduce complexity by 75\% without sacrificing localization performance by going to $(3,6)$, or
improve localization performance by 50\% without increasing complexity by going to $(18,6)$, both 
representing a significant improvement. 


To understand why grouping pilots is a good idea, in Fig.~\ref{response} 
we plot the RSS at the AP as reflected by the RIS from different starting locations in the $(x,y)$ plane.
The sub-figure at the $j$-th column and $i$-th row is
obtained under the RIS configuration produced by the LSTM for the $j$-th pilot
at the $i$-th stage, when the UE, RIS, and BS are located at the red,
black, and yellow points, respectively.
Cyan lines represent the propagation paths reflecting once, while magenta lines
depict paths reflecting twice.  For visual clarity, paths corresponding to two
reflections are included only in the first sub-figure, and paths with more than
three reflections are not shown.

We observe in Fig.~\ref{response} that the RIS appears to be probing different directions within the same
stage to explore multiple paths.  
This interpretation of the designed RIS configuration underscores the efficacy of the proposed
method in grouping multiple pilots to explore multiple paths within each stage.
Based on the observation from Fig.~\ref{accuracy}, Fig. \ref{response} and Fig.~\ref{time}, it is evident that the proposed method yields interpretable results and effectively reduces both training and inference complexity, while maintaining good localization performance.
 
\section{Conclusion}

This letter proposes a method that employs LSTM networks to tackle the uplink
RIS-assisted localization problem in narrowband SISO systems, specifically
focusing on multipath propagation scenarios. By grouping pilot signals and
utilizing grouped pilots as input for the LSTM, the proposed method can reduce
training and inference complexity by limiting the number of concatenated active
sensing units.  We find that the RIS configuration can be dynamically
adjusted throughout the LSTM learning process to explore different
reflection paths within a single stage, thus enabling superior localization 
accuracy at low complexity. 

\bibliographystyle{IEEEtran}
\bibliography{IEEEabrv,reference}

\vfill

\end{document}